# Drift Correction of Scan Images by Snapshot Referencing

*Zac Thollar*[1], *Kanto Maeda*[1], *Tetsuya Kubota*[1], *Taka-aki Yano*[2], *Qiwen Tan*[1], *Takumi Sannomiya*[1*]

AUTHOR ADDRESS

[1] Department of Materials Science and Engineering, School of Materials and Chemical Technology, Institute of Science Tokyo, 4259 Nagatsuta, Midori-ku, Yokohama 226-8501 Japan

[2] Institute of Post-LED Photonics, Tokushima University, 2-1 Minami-Josanjima, Tokushima, 770-8506, Japan

*sannomiya@mct.isct.ac.jp

## ABSTRACT

Reliable quantitative analysis in scanning (transmission) electron microscopy (S(T)EM) is often hindered by image drift during long-duration spectral mapping for elemental analysis or for various material functions. We here present snapshot-referencing (SSR) drift correction, a retrospective approach to eliminate spatial distortion based on the temporal nature of the scanning process. A continuous drift vector for every pixel is calculated for a normalized time-field of the scan pattern (e.g., serpentine or raster) utilizing a high-signal, fast-scan "snapshot" as a drift-free reference to guide the correction of simultaneously acquired analytical maps. To describe the drift, we employed Bezier basis functions to model smooth thermal or mechanical drifts and piece-wise linear basis for high-frequency "spiky" shifts such as those caused by charging. We demonstrate the efficacy of this approach on experimental cathodoluminescence (CL) datasets, showing that it effectively restores spatial integrity to hyperspectral data cubes without the need for specialized hardware. This flexible, software-based solution is broadly applicable to any probe-based analytical technique where a fast imaging signal can be recorded alongside slow spectroscopic data.

# Introduction

In scanning (transmission) electron microscopy (S(T)EM), spatially resolved analytical techniques, such as energy-dispersive X-ray spectroscopy (EDS), electron energy-loss spectroscopy (EELS), and cathodoluminescence (CL), play a central role in the investigation of materials and devices.[1–4] These methods enable nanoscale mapping of composition, electronic structure, and optical properties and are therefore indispensable in materials science, nanotechnology, and semiconductor research. [5] However, acquiring high-quality analytical maps typically requires long dwell times and extended total acquisition times, making the measurements highly susceptible to image drift. Image drift during long scans arises from a combination of thermal instability, mechanical relaxation, and environmental fluctuations in and around the microscope. [6] Due to the nature of the electron beam, sample charging also introduces high-frequency stochastic image shifts. Even small drifts can severely degrade analytical maps, leading to distortion, blurring, and misregistration between structural and spectroscopic information. As a result, reliable drift correction is essential for quantitative and high-resolution analysis. Conventional solutions often involve real-time hardware (either stage or beam) tracking, the acquisition of multiple images with different scan angles, or 4D-STEM ptychography, which can be hardware-intensive. [6–9] Software-based distortion or drift correction methods have also been proposed; these typically rely on multiple image acquisition or on specific features of the specimen. [10,11] The common problem is that a direct distortion-free (i.e. the "ground truth") is rarely available in most practical cases.

In analytical spectral mapping, such as EELS, EDS or CL, slow scan data acquisition is usually performed simultaneously with the imaging of high-intensity signals such as secondary electron (SE), bright-field (BF), annular dark-field (ADF) or panchromatic images. Because these signals have much higher count rates than spectroscopic signals, they can be acquired rapidly as snapshots with a high signal-to-noise ratio and with minimal sensitivity to image drift. This contrast in acquisition speed and signal strength suggests that the fast-scan images could serve as a robust snapshot reference for correcting distortions in simultaneously acquired analytical maps. Thus, reliable "solutions" may exist for the drift problems in slow-scan analytical mapping.

In this work, we propose an effective drift correction method that uses a high-signal, fast-scan snapshot as a reference to correct the image distortion of long-acquisition spectral maps (Fig. 1a). By quantifying the distortion between the fast reference image and the corresponding image acquired simultaneously with the spectral map, the drift in the slow-scan data can be estimated and compensated. Unlike conventional hardware-based real-time drift correction methods, the proposed snapshot-referencing (SSR) approach does not rely on specific microscope hardware and can be applied retrospectively, independent of the timing of drift correction during acquisition. This makes the method simple, flexible, and broadly applicable to a wide range of probe-based analytical techniques. We demonstrate the proposed SSR approach using simulated images as well as experimentally obtained CL spectral mapping data with image distortions originating from various types of drift.

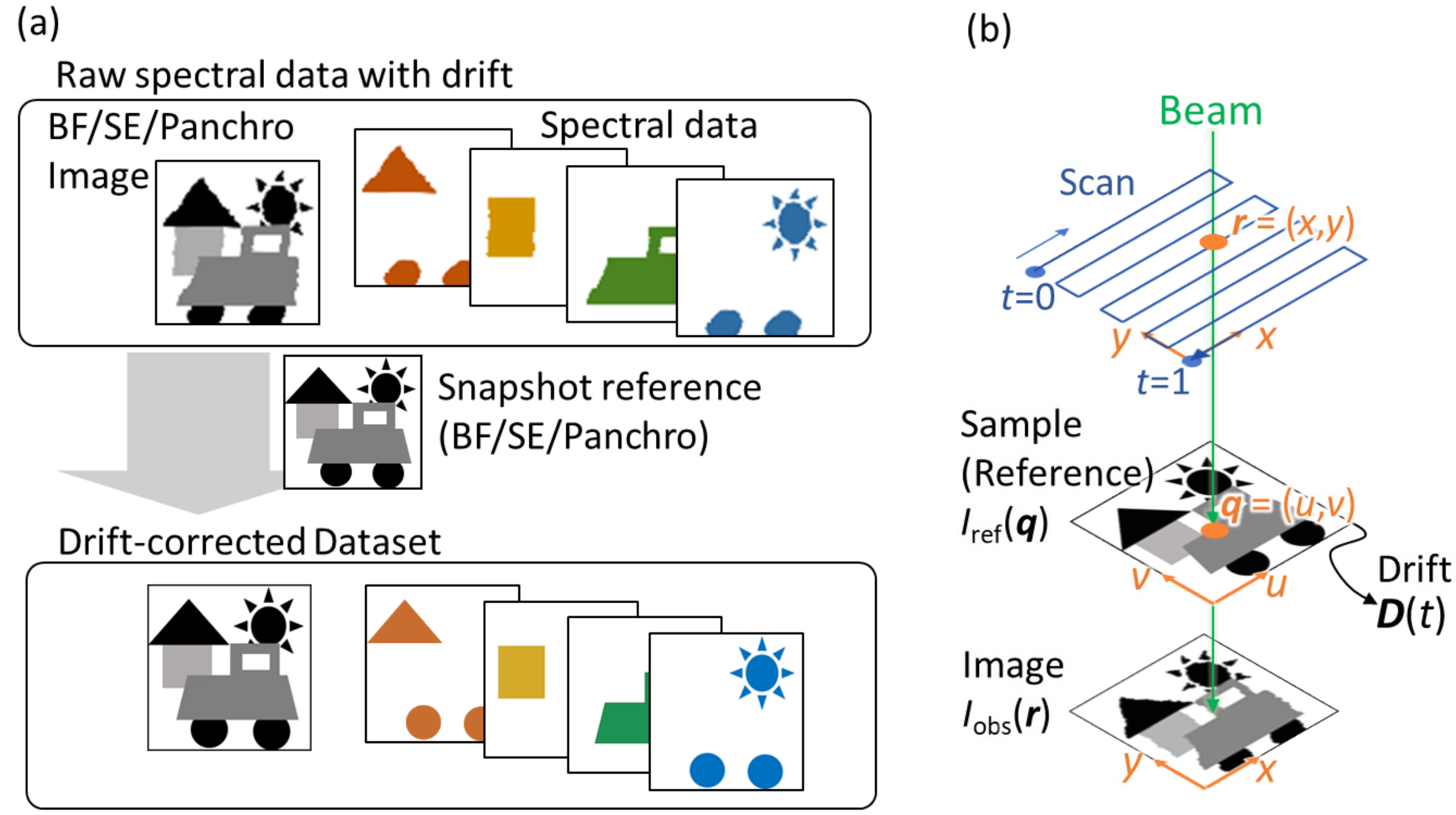

**Figure 1. Snapshot-referencing (SSR) drift correction.** (a) Concept of SSR. A fast scan snapshot reference image with high signal intensities is used to correct the image distortion of the spectral data set. (b) Description of the drift during the serpentine beam scan and the definition of the parameters.

# Methods

### *Theory*

The sample drift situation is schematically illustrated in Fig 1b. We assume that the distortion is caused by the temporal drift of the scanning area with respect to the sample position and we correct the drifted image by estimating a time-dependent drift function $\boldsymbol{D}(t)$. Since the drift happens over time, we first define a normalized time $t(x, y) \in [0, 1]$, scaled by the total scan duration. This time-field directly corresponds to each scan pixel position $\boldsymbol{r} = (x, y)$ in the scan, determined by the specific scanning pattern, i.e. raster or serpentine. Every pixel in the raw image at $\boldsymbol{r} = (x, y)$ is then mapped to its corresponding warped position $\boldsymbol{q} = (u, v) = \boldsymbol{r} + \boldsymbol{D}$ in the reference (or sample) coordinate space (Fig. 1b). Ideally, the observed image intensity $I_{obs}$ at a given beam position directly corresponds to the sample information $I_{\text{ref}}$ of the warped position.

The objective is to estimate unknown drift function $\boldsymbol{D}$ to minimize the difference between $I_{\text{obs}}(\boldsymbol{r})$ and $I_{\text{ref}}(\boldsymbol{r} + \boldsymbol{D})$. To find the optimal solution for $\boldsymbol{D}$, we formulate an energy functional $E[\boldsymbol{D}]$ of the system to be minimized. Since the problem is ill-posed, appropriate regularization is required. To guarantee the temporal continuity of the drift, we introduce the $H^1$ semi-norm (Dirichlet energy) weighted by $\lambda$, and define the energy functional as follows:

$$E[\boldsymbol{D}] = E_{\text{dat}}[\boldsymbol{D}] + \lambda E_{\text{reg}}[\boldsymbol{D}] \ . \tag{1}$$

Here, $E_{\text{dat}}$ is the squared error between the observed and reference images, representing the potential energy:

$$E_{\text{dat}}[\boldsymbol{D}] = \int (I_{\text{ref}}(\boldsymbol{r} + \boldsymbol{D}) - I_{\text{obs}}(\boldsymbol{r}))^2 d\boldsymbol{r} \ . \tag{2}$$

And $E_{\text{reg}}$ is the regularization term, corresponding to the kinetic energy of the drift motion:

$$E_{\text{reg}}[\boldsymbol{D}] = \int |\frac{d\boldsymbol{D}}{dt}|^2 dt \ . \tag{3}$$

Since the optimal $\boldsymbol{D}$ minimizes the functional $E[\boldsymbol{D}]$, i.e. $\delta E = 0$, the following Euler–Lagrange equation can be obtained after inserting Eqs.(2) and (3) into Eq. (1) and taking its derivative:

$$\left(I_{\mathrm{ref}}(\boldsymbol{r}+\boldsymbol{D}) - I_{\mathrm{obs}}(\boldsymbol{r})\right) \nabla I_{\mathrm{ref}}(\boldsymbol{r}+\boldsymbol{D}) - \lambda \frac{d^2\boldsymbol{D}}{dt^2} = 0 \ . \tag{4}$$

This equation represents the balance between the external force in the first term, which tries to align the images, and the tension of the temporal drift path in the second term, which tries to smooth the drift function, similar to the restoring force of a tensioned string. Therefore, $\lambda$ is a parameter that adjusts the trade-off between fidelity to the data and smoothness of the drift. Such regularization is particularly important when the image contrast is flat, for instance in empty regions.

***Implementation***

Since the drift correction corresponds to finding the drift $\boldsymbol{D}$ that minimizes the total energy $E$ of Eq.(1), we now implement this optimization problem using the practical pixelized images, i.e. slow scan drifted image $I_{\mathrm{obs}}$ and fast scan snapshot reference $I_{\mathrm{ref}}$. While $E_{\mathrm{dat}}$ in Eq. (2) is simply the intensity difference of these two images, $E_{\mathrm{reg}}$ in Eq. (3) corresponds to the power of the drift position change per discretized time. Therefore, we can define a loss function $\mathcal{L}$ to be minimized for the practical discretized dataset corresponding to Eq.(1):

$$\mathcal{L} = \sum_i \left(I_{\mathrm{ref}}(\boldsymbol{r}_i) - I_{\mathrm{obs}}(\boldsymbol{q}_i)\right)^2 + \lambda \sum_i \frac{|\boldsymbol{D}_{i+1} - \boldsymbol{D}_i|^2}{\Delta t}, \tag{5}$$

where $i$ is the serial index of the scan position, each corresponding to the time index. Now we model the time-dependent drift vector $\boldsymbol{D}(t)$ as a sum of two components to capture both low-frequency trends, expressed by a Bezier basis $\boldsymbol{D}^{\mathrm{bez}}(t)$, and high-frequency local jitters, expressed by a piece-wise linear shift $\boldsymbol{D}^{\mathrm{lin}}(t)$:

$$\boldsymbol{D}(t) = \boldsymbol{D}^{\mathrm{bez}}(t) + \boldsymbol{D}^{\mathrm{lin}}(t) . \quad (6)$$

The low-frequency Bezier component is expressed as $\boldsymbol{D}^{\mathrm{bez}}(t) = \sum_{j=0}^{n} \boldsymbol{P}_j \, B_{j,n}(t)$, where $B_{j,n}(t)$ are Bernstein basis polynomials and $\boldsymbol{P}_j$ are control points optimized by the solver. The high-frequency piece-wise linear function is defined as $\boldsymbol{D}^{\mathrm{lin}}(t) = \mathrm{Interp}(\mathbf{N}_0, \mathbf{N}_1 \ldots, \mathbf{N}_k, t)$, where $\mathbf{N}_k$ are discrete drift values at specific timestamps and the Interp(:) function linearly interpolates these drift values at time *t*. To include this drift $\boldsymbol{D}$ in the resultant image, we map every pixel in the observed image $\boldsymbol{r} = (x, y)$ to its corresponding warped position $\boldsymbol{q} = (u, v)$ in the reference (or true sample) image space, as illustrated in Fig. 1b. $I_{\mathrm{ref}}$ should be interpolated to describe sub-pixel drifts. In practice, since the reference and observed images are obtained separately under different conditions, this mapping can include a global affine transform (zoom/shift) in order to adjust the scale and position alignment of the "reference" image. Thus, the warped position is redefined as:

$$\boldsymbol{q} = \mathbf{S} \; \boldsymbol{r} + \Delta\boldsymbol{r} + \boldsymbol{D} \; . \quad (7)$$

The first term $\mathbf{S}$ and the second term $\Delta\boldsymbol{r}$ correspond to the diagonal zoom matrix and the shift, respectively.

To accommodate large drifts as well as to decouple all the parameters, an iterative alternating minimization scheme is applied where the global affine transformations and local non-linear distortions are performed alternately during the algorithm cycle, which allows gradually increasing the model complexity to ensure stable convergence. To show the robustness of the proposed approach, the same parameter setting is used for all the presented data in this study: 50 Bezier bases, 50 linear nodes, $\lambda = 10^{-2}$, and 20 iterations. These numbers were determined through several trials using simulation images of 75 x 75 pixels, which we discuss later, though they are not fully optimized. The base and node numbers should be increased as the drift becomes severer and the pixel number increases. The optimal penalty weight $\lambda$ would also depend on the portion of flat areas. We employed the sequential quadratic programming algorithm for the optimization. While we present a specific implementation of the SSR drift correction framework, the architecture permits further optimization other than simple parameter settings. For instance, the Bezier basis functions used to ensure global smoothness may be omitted, provided a sufficient density of nodes and optimized nodal positioning are utilized in the piece-wise linear basis. Furthermore, other specialized basis functions (e.g., B-splines or wavelets) may be integrated depending on the specific noise characteristics. Regarding regularization, we observed that an $L^2$ norm penalty on the parameter sets is equally effective in constraining the magnitude of the basis coefficients, thereby preventing over-fitting and ensuring physically plausible drift estimates. Also, any type of scan pattern including multiple scans can be readily implemented by defining the scan position function $\boldsymbol{r}(t)$ accordingly.

### *Experimental spectral mapping: cathodoluminescence*

For the experimental spectral mapping, we performed cathodoluminescence (CL) mapping. Modified STEM (2100F and 2000FX, JEOL, Japan) instruments equipped with a CL detection setup were utilized. [4,12] In the CL setup, a parabolic mirror situated at the sample position collimates the light emission from the sample. CL signals are detected by integrating the entire emission angle covered by the parabolic mirror. The CL spectral maps are acquired by dispersing the light with an optical grating and detected in a CCD detector, resulting in 3D information consisting of the *x*-*y* electron beam scan dimensions and the dispersed wavelength. The electron beam is scanned in a serpentine manner for the spectroscopic mapping, starting from the top-left to the bottom in the presented images. For the reference images, either a STEM BF image (for Experimental data I and II) or a panchromatic CL image (for Experimental data III) acquired by a fast scan is used and compared with the simultaneously acquired BF image or integrated CL (panchromatic) image obtained during the long-duration CL spectral mapping. These data were used for the correction. A CL spectral mapping typically requires more than 10 minutes of acquisition, while the snapshot reference BF or panchromatic CL image can be acquired within a few seconds (for BF) or few minutes (for panchromatic CL) with much higher pixel resolution. For the snapshot reference panchromatic image, a separate photon counting detector without wavelength dispersion was used.

## Results and Discussion

### *Simulation image*

We first apply the SSR drift correction to simulated images, as shown in Fig. 2. Here, a distorted image is artificially generated (Fig. 2a) from a reference image (Fig. 2b) by introducing random low-frequency and stochastic high-frequency drifts during a serpentine scan. The drift function estimated using the proposed SSR algorithm is shown in Fig. 2c, and for comparison, the introduced artificial drift in the simulation is shown in Fig. 2d. The drift-corrected image in Fig. 2e shows that the image distortions at both high and low frequencies are effectively corrected, confirming the applicability of the method. While the image distortion seems properly corrected, the estimated drift function in Fig. 2c shows different background (seemingly linear) ramps in both the X and Y directions compared to the originally introduced drift data in Fig. 2d. This difference originates from the affine transform process (magnification and global shift adjustment) described in Eq. (7) because the magnification adjustment can compensate for the linear drift in the scan image.

To evaluate the difference from the reference image, the squared residual and structural similarity index measure (SSIM) are shown in Figs. 2f and 2g. Both assessments show that the corrected image closely matches the reference, except for the edge areas (e.g., around the ear of the deer) where the original distorted image lacks data because “the sample” drifted out of the frame. Other evaluation details, such as loss function convergence and drift component decomposition, are shown in the Supplementary Material (SM). Thus, the simulation clearly demonstrates the validity of the proposed SSR algorithm. We also note that the drift-corrected image is slightly smoothed due to the interpolation process in the algorithm.

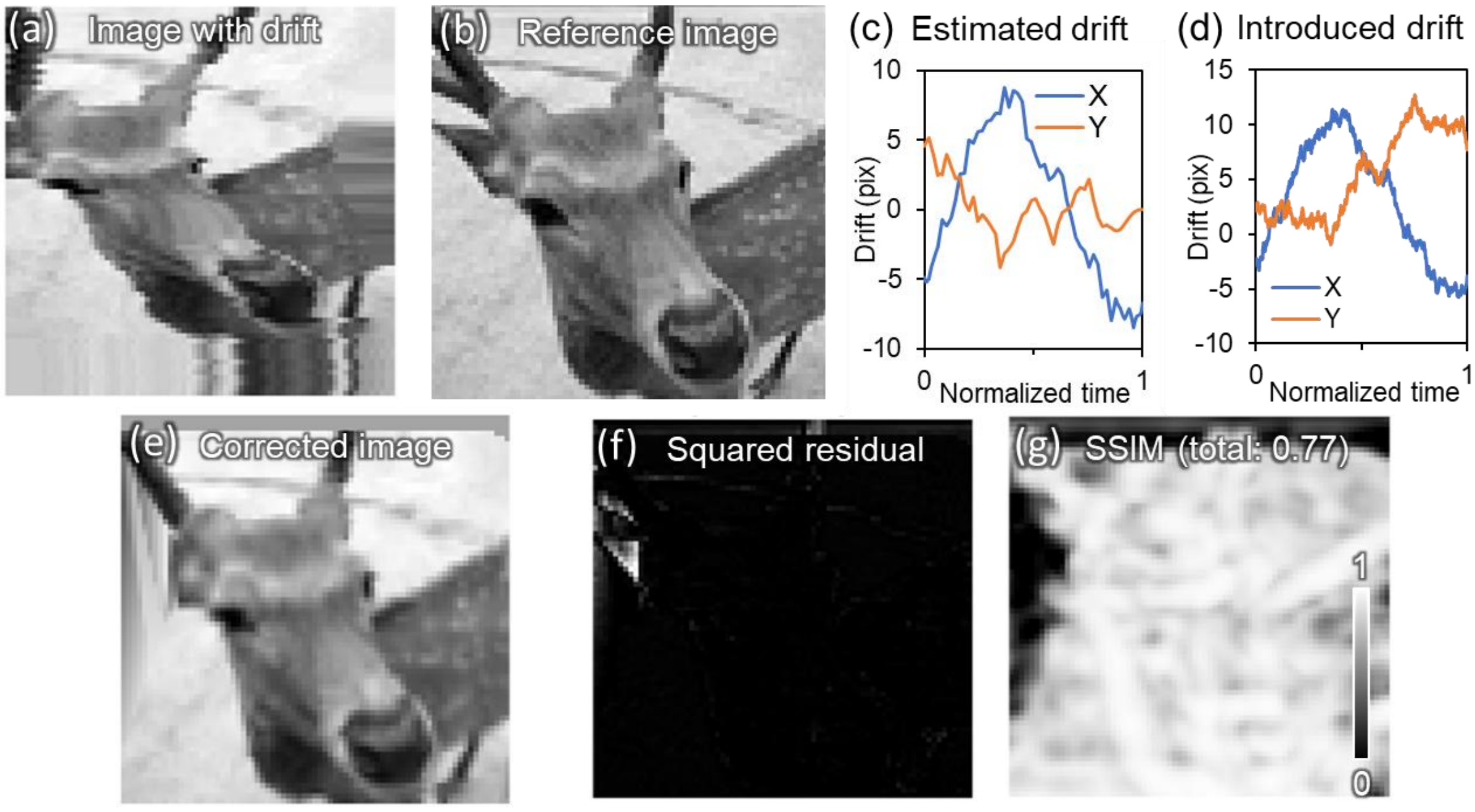

**Figure 2. Simulated drift correction.** (a) Image with artificially introduced drifts. (b) Reference image. (c) Estimated time-dependent drifts in the X and Y directions. (d) Originally introduced drift in the simulation for comparison. (e) Drift-corrected image obtained from the image in panel

a by referencing the image in panel b. (f) Squared residual and (g) SSIM maps between the reference and corrected images. All the images are 75 × 75 pixels.

### *Experimental data I*

We now apply the SSR method to experimentally obtained hyperspectral CL data. CL spectral mapping typically requires tens of minutes for data acquisition and may be subject to sample drift due to instrumental instability, sample movement, or environmental changes such as temperature fluctuations. Here, we show a CL mapping example of plasmonic silver (Ag) nanoparticles exhibiting optical near fields, which can be visualized by CL.

Figure 3 shows the original and drift-corrected CL data along with a reference image and the drift evaluation results. The original CL dataset in Figs. 3a-c exhibits a low frequency distortion, possibly due to temperature fluctuations, which is clearly visible when compared with the reference snapshot image in Fig. 3d. The drift is significant, especially in the middle of the image. The SSR-corrected results and subsequent evaluations are shown in Figs. 3e-j. The corrected BF map in Fig. 3f well matches the reference image in Fig. 3d. The correspondingly corrected CL datasets in Figs. 3g and 3h also show reasonable dipolar patterns for each particle. The image correspondence between the reference (Fig. 3d) and corrected BF (Fig. 3f) images is evaluated using the squared residual and SSIM in Figs. 3i and 3j, respectively. Both evaluations show strong correspondence between the two images. We note that the distortion correction parameters can be optimized for each dataset, while in this demonstration the same parameter set is used in the previous simulation and the following example in the next section.

The time-dependent drift function shows that the drift is significant in the middle of the scan and is composed mostly of slow (low frequency) components. The high frequency components are mostly within a single pixel range. The decomposed Bezier and piecewise-linear drifts are separately plotted in the SM. As expected, the Bezier component plays the primary role in the drift correction.

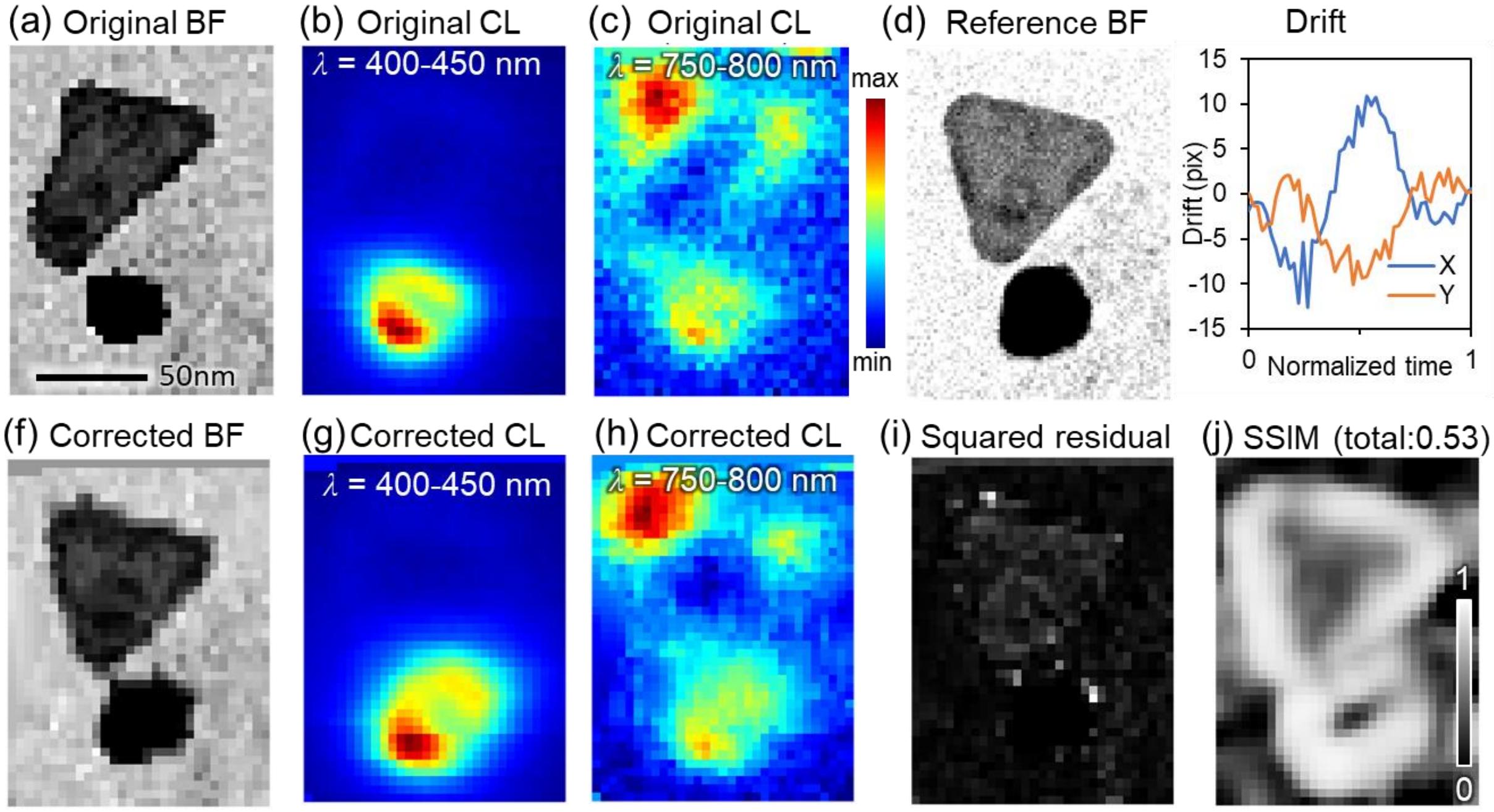

**Figure 3. CL mapping result of Ag nanoparticles with low frequency drifts.** (a) BF image obtained during the spectral mapping, and (b, c) representative spectral maps respectively at the wavelengths of 400 – 450 nm and 750 – 800 nm with some low frequency drifts. (d) Snapshot reference BF image to correct the BF image of panel a. (e) Estimated time-dependent drifts in the X and Y directions. (f) Drift-corrected BF image from panel a. (g, h) Drift-corrected spectral maps obtained from images in panels b and c respectively. (i) Squared residual and (j) SSIM maps between the reference and corrected images. The hyperspectral CL mapping was acquired with the scan size of $31 \times 41$ pixels and the total acquisition time of ~15 min.

### *Experimental data II*

To further demonstrate the effectiveness of SSR drift correction, we present another CL example exhibiting high-frequency stochastic drifts. The sample is an oxide particle, which is susceptible to charging by the slow-scan electron beam, potentially resulting in abrupt image shifts. Figure 4 shows the original and corrected CL datasets along with the reference image and evaluation results. Because of the weaker CL signal compared to the plasmonic nanoparticles shown in Fig. 3, the acquisition time is longer: ~35 min for a scan of 41 x 41 pixels. We emphasize again that the same correction parameter set was used also for this example as in the previous two examples.

The original slow-scan data in Figs. 4a-c exhibit spiky drifts, which are clearly visible in both the BF and CL maps. The SSR drift correction effectively compensates for these artifacts, producing a BF image (Fig. 4f), almost identical to the reference (Fig. 4d), and smooth CL maps (Figs. 4g, h). The time-dependent drift function in Fig. 4e shows the spiky drift traces, as expected. The squared residual and SSIM maps comparing the corrected and reference BF images both confirm a strong match. The decomposed drift maps for the Bezier and piecewise-linear drift components show a dominant contribution from the piecewise-linear nodes, which effectively model the spiky high frequency drifts.

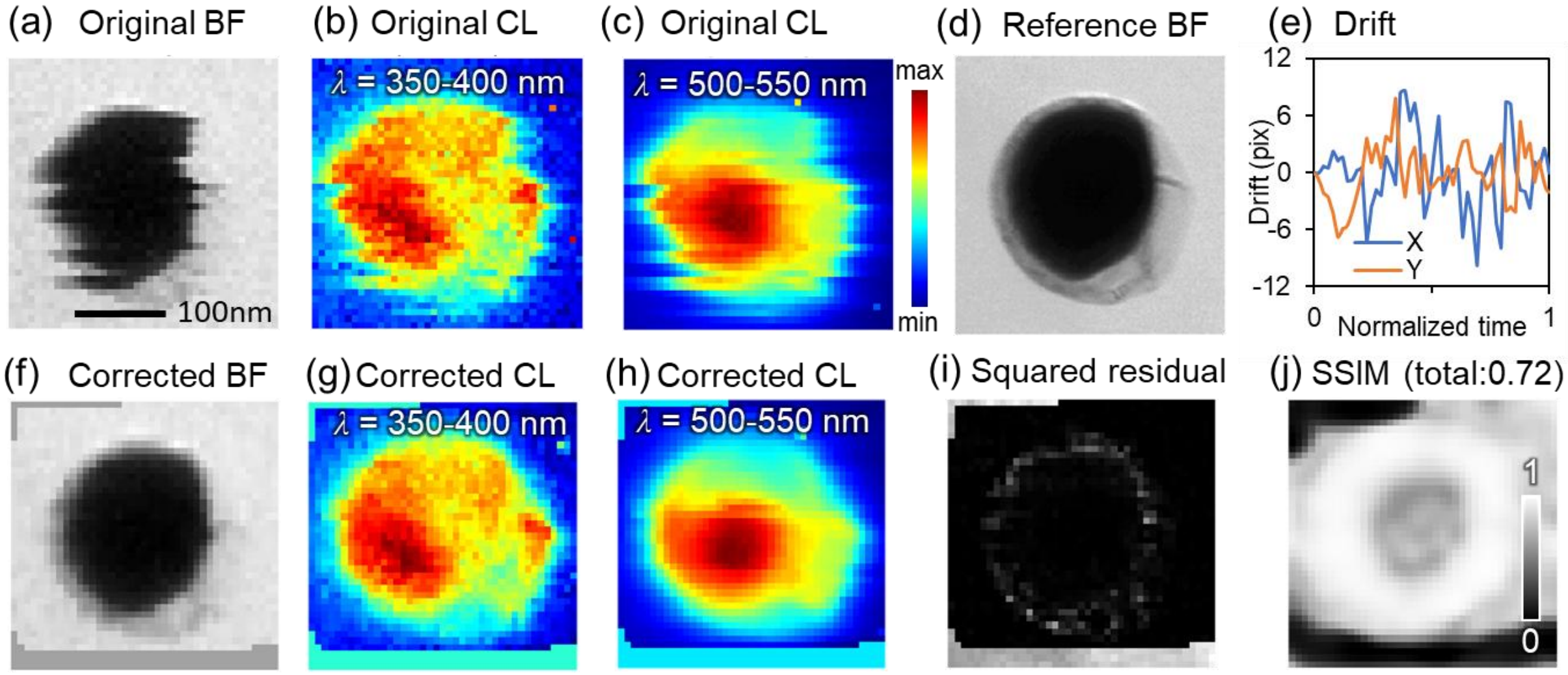

**Figure 4. CL mapping result of a $TiO_2$ nanoparticle with high frequency spiky drifts.** (a) BF image obtained during the spectral mapping, and (b, c) representative spectral maps respectively at the wavelengths of 350 – 400 nm and 500 – 550 nm with spiky drifts. (d) Snapshot reference BF image to correct the BF image of panel a. (e) Estimated time-dependent drifts in X and Y directions. (f) Drift-corrected BF image from panel a. (g, h) Drift-corrected spectral maps from panels b and c respectively. (i) Squared residual and (j) SSIM maps between the reference and corrected images. The hyperspectral CL mapping was acquired with the scan size of $41 \times 41$ pixels and the total acquisition time of ~35 min.

***Experimental data III***

To verify the applicability of the method, we also used a panchromatic CL image as a reference. This image can be obtained using a single channel detector to collect CL signals of all the wavelengths, resulting in a much higher signal-to-noise ratio and allowing for faster acquisition time, compared to dispersive spectral detection. In the slow-scan spectral map, the corresponding panchromatic image is obtained by integrating the signal over all the wavelengths. Figure 5 shows a demonstration of the drift correction for such a spectral mapping containing several image drifts. A cluster of nanodiamond particles with nitrogen-vacancy emission centers was used as a specimen. As shown in the figure panels, the spiky shift and gradual image distortions are effectively corrected by the SSR correction algorithm. Additional correction data are also available in the SM.

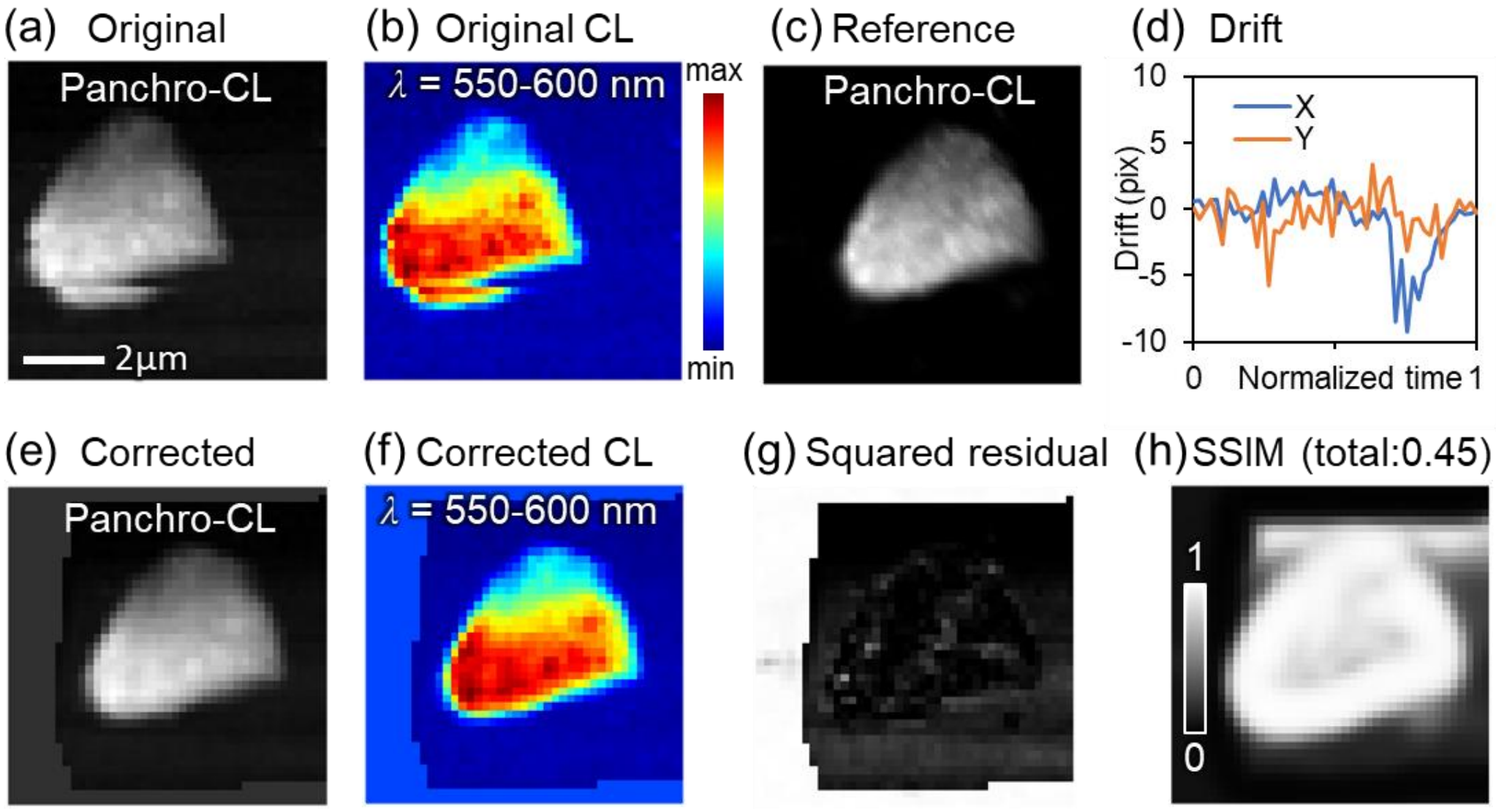

**Figure 5. CL mapping result of a nanodiamond cluster.** (a) Slow-scan panchromatic CL image obtained by integrating the spectral map, and (b) representative spectral map at the wavelengths of 550 – 600 nm. (c) Fast-scan snapshot reference panchromatic CL image to correct the image of panel a obtained during the spectral mapping. (d) Estimated time-dependent drifts in X and Y directions. (e) Drift-corrected panchromatic CL image from panel a. (f) Drift-corrected spectral maps from panels b. (g) Squared residual and (h) SSIM maps between the reference and corrected images. The hyperspectral CL mapping was acquired with the scan size of 41 × 41 pixels and the total acquisition time of ~15 min.

## Conclusion

In this work, we have demonstrated the snapshot-referencing (SSR) drift correction method as a robust solution for improving the spatial accuracy of analytical S(T)EM mapping. By treating drift as a continuous function of acquisition time rather than a simple global shift, the SSR algorithm successfully compensates for complex, non-linear distortions that accumulate during long dwell times. Here, we utilized Bezier and linear basis functions to cover drifts of various origins ranging from slow mechanical relaxation to sudden charging-induced jumps. It should be noted that any basis function can be applied but the key insight here is the introduction of a time-dependent drift function and the utilization of a reference image. Regularization of the drift function suppresses unnatural or unnecessary image deformations. Once the high signal data is corrected, every spectral signature is remapped to its "true position" by warping the entire 3D data cube. As a post-processing technique, this approach requires no specialized real-time correction hardware, making it accessible for a wide range of microscope systems and varied experimental setups. The presented experimental results confirm that the SSR method restores structural fidelity and minimizes misregistration between spectroscopic data and imaging signals. As the demand for higher spatial resolution in analytical microscopy grows, the proposed SSR approach provides a reliable framework for handy yet quantitative mapping even in the presence of environmental or sample instabilities. To further expand this approach, the constraint of image-matching with the "reference" in the algorithm could be more flexibly modified to include factors such as smoothness of the reconstructed image or feature matching. In the iteration process, parameter-adaptive approaches would also improve the convergence.

## Funding Sources

This work is supported by JSPS Kakenhi (JP24H00400), JST CREST (JPMJCR25I3), and the Mitsubishi Foundation (202510021).

## Corresponding author

Takumi Sannomiya : sannomiya@mct.isct.ac.jp

## Supplementary material

Loss convergence data during the iteration and decomposed drift components of each correction data.

## Competing interests

The authors declare no competing interests

# Supplementary Material for:

# Drift Correction of Scan Images by Snapshot Referencing

*Zac Thollar*[1], *Kanto Maeda*[1], *Tetsuya Kubota*[1], *Taka-aki Yano*[2], *Qiwen Tan*[1], *Takumi Sannomiya*[1*]

AUTHOR ADDRESS

[1] Department of Materials Science and Engineering, School of Materials and Chemical Technology, Institute of Science Tokyo, 4259 Nagatsuta, Midori-ku, Yokohama 226-8501 Japan

[2] Institute of Post-LED Photonics, Tokushima University, 2-1 Minami-Josanjima, Tokushima, 770-8506, Japan

*sannomiya@mct.isct.ac.jp

## S1. Additional analysis results for the simulated data

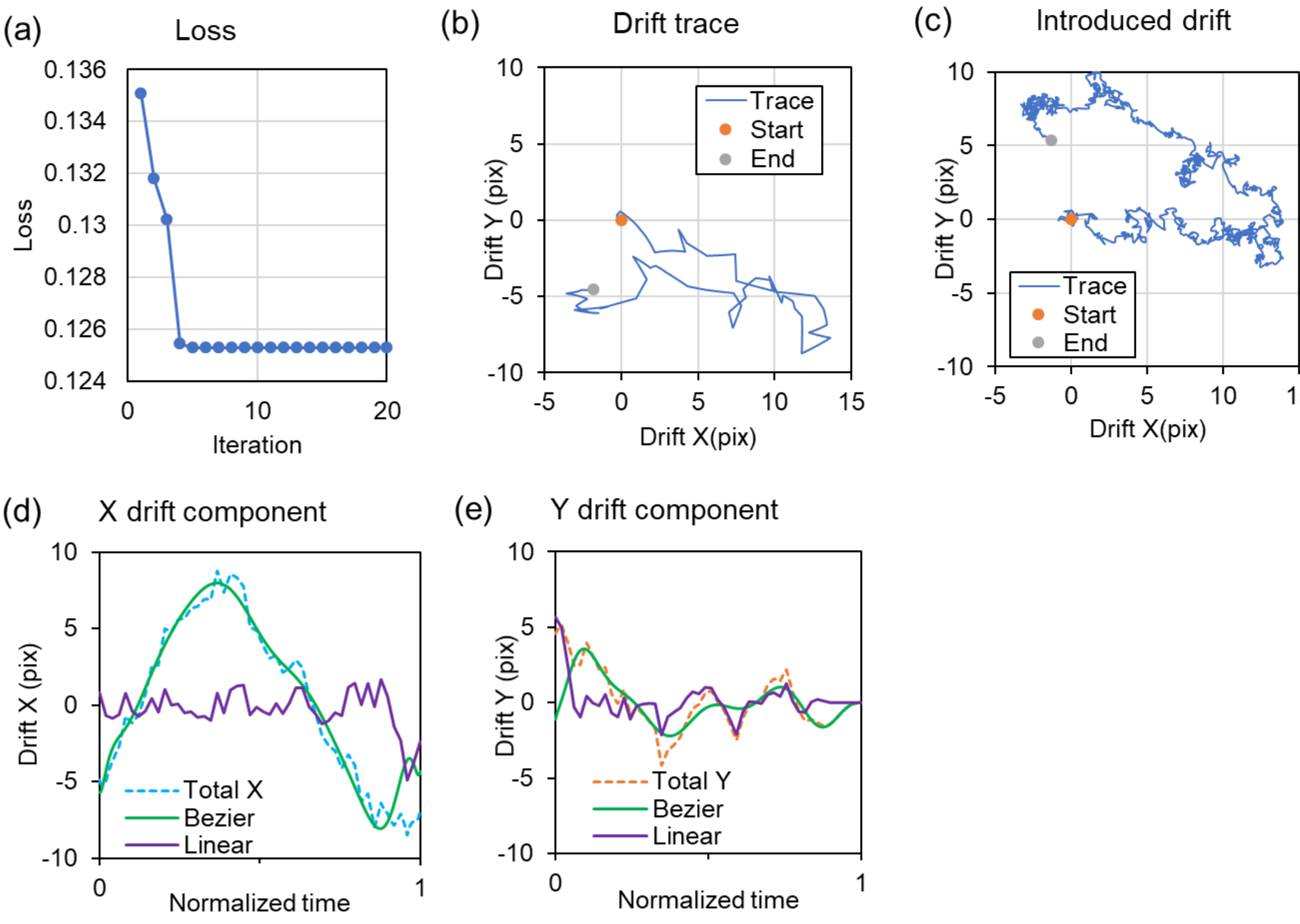

**Figure S1. Additional analysis results for the simulated data.** (a) Loss function change by iteration. (b) Drift trace plotted in the *x*-*y* plane. (c) Trace plot of the originally introduced drift in the simulation, for comparison. (d,e) Bezier and piece-wise-linear drift components chronologically plotted for (c) *x* and (d) *y*. The start positions of the trace plots in panels b and c are offset to the origin.

## S2. Additional analysis results for the Ag nanoparticle data

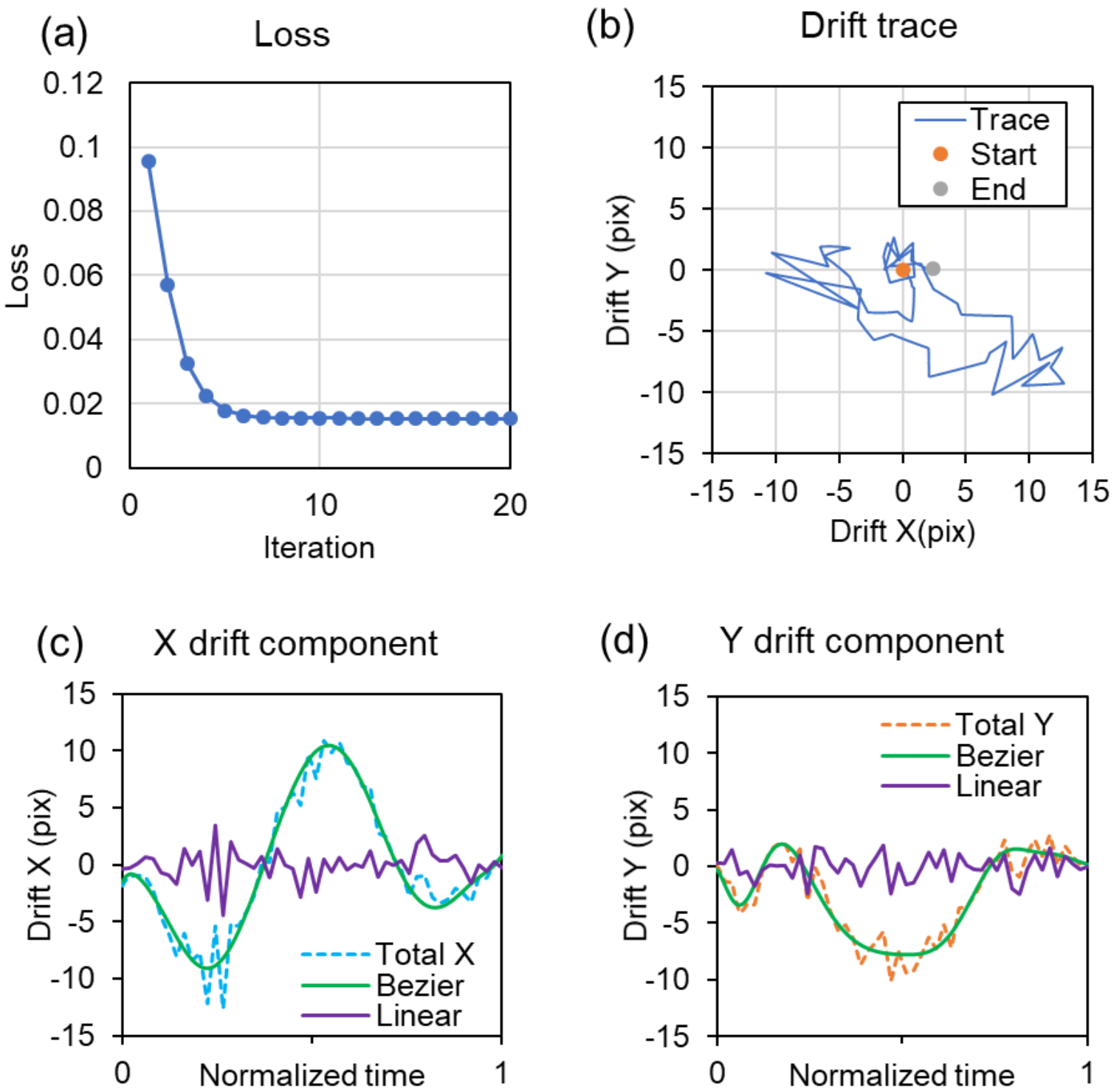

**Figure S2. Additional analysis results for the Ag nanoparticle data.** (a) Loss function change by iteration. (b) Drift trace plotted in the $x$-$y$ plane. (c,d) Bezier and piece-wise-linear drift components chronologically plotted for (c) $x$ and (d) $y$. The start position of the trace plot in panel b is offset to the origin.

## S3. Additional analysis results for the $TiO_2$ nanoparticle data

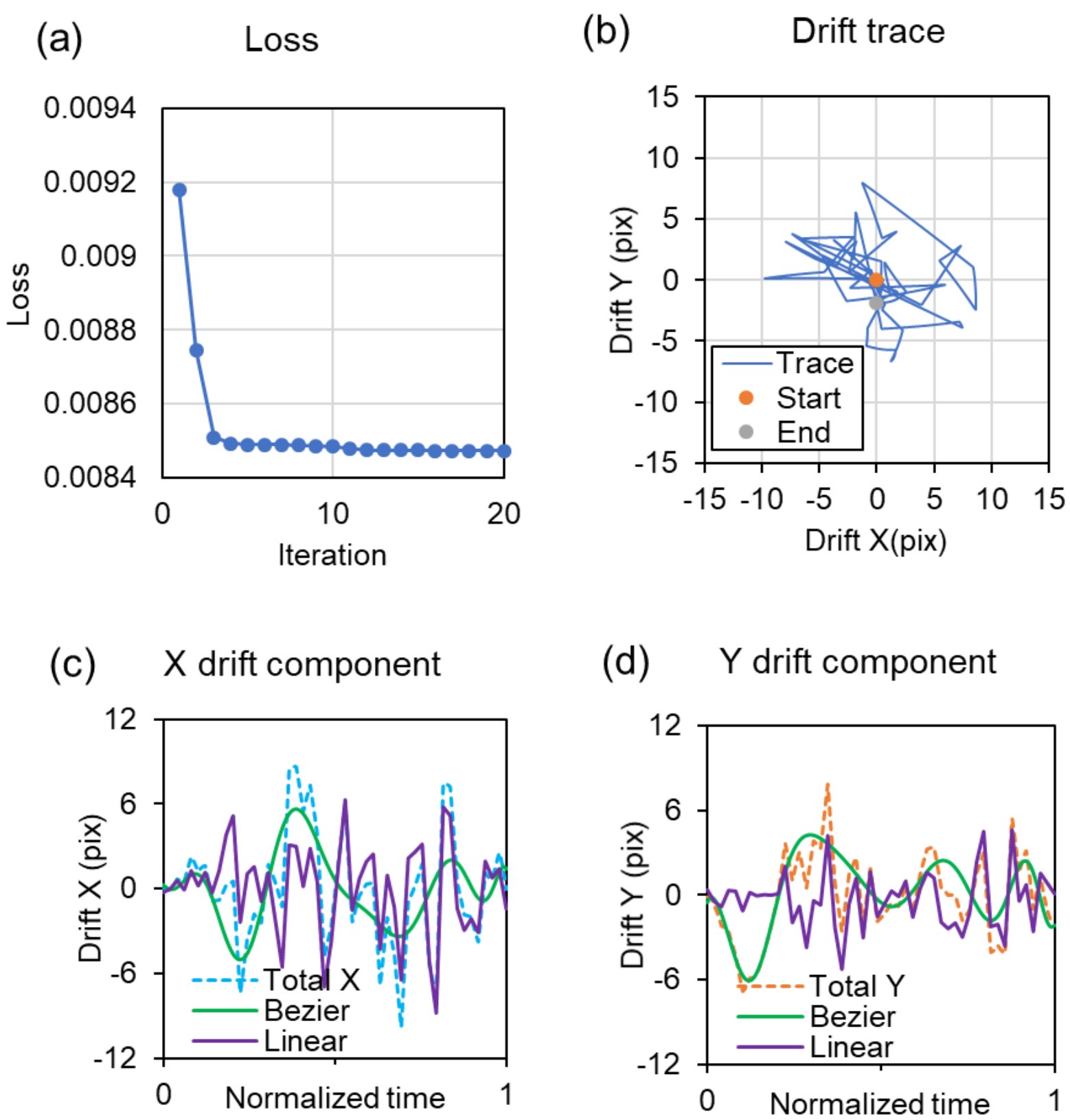

**Figure S3. Additional analysis results for the $TiO_2$ nanoparticle data.** (a) Loss function change by iteration. (b) Drift trace plotted in the *x*-*y* plane. (c,d) Bezier and piece-wise-linear drift components chronologically plotted for (c) *x* and (d) *y*. The start position of the trace plot in panel b is offset to the origin.

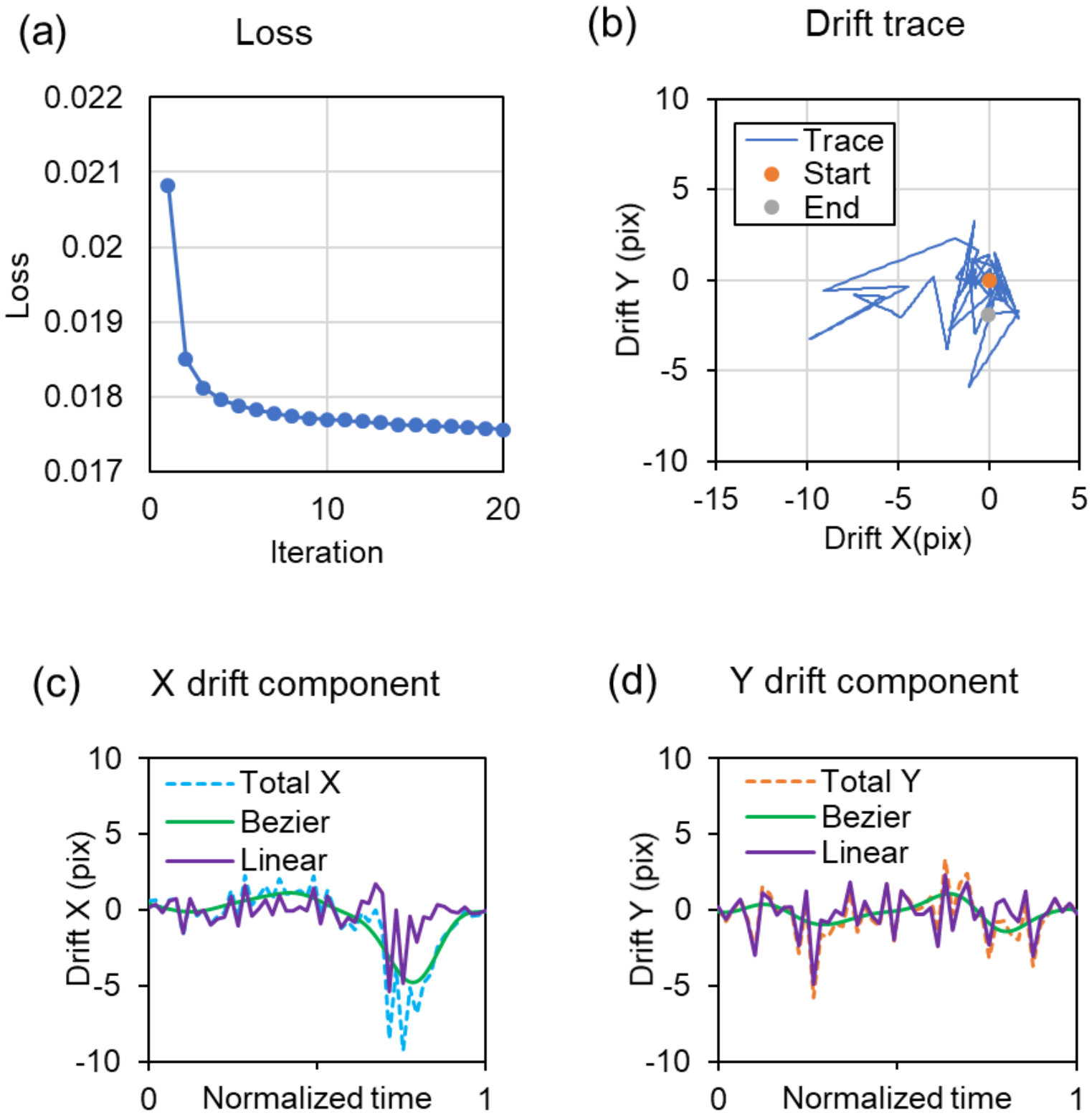

**Figure S4. Additional analysis results for the nanodiamond cluster data.** (a) Loss function change by iteration. (b) Drift trace plotted in the $x$-$y$ plane. (c,d) Bezier and piece-wise-linear drift components chronologically plotted for (c) $x$ and (d) $y$. The start position of the trace plot in panel b is offset to the origin.